# Bottom-up fabrication of highly ordered metal nanostructures by hierarchical self-assembly

*Denise J. Erb*†, Kai Schlage, Ralf Röhlsberger*

Deutsches Elektronen-Synchrotron DESY, Notkestraße 85, 22607 Hamburg, Germany

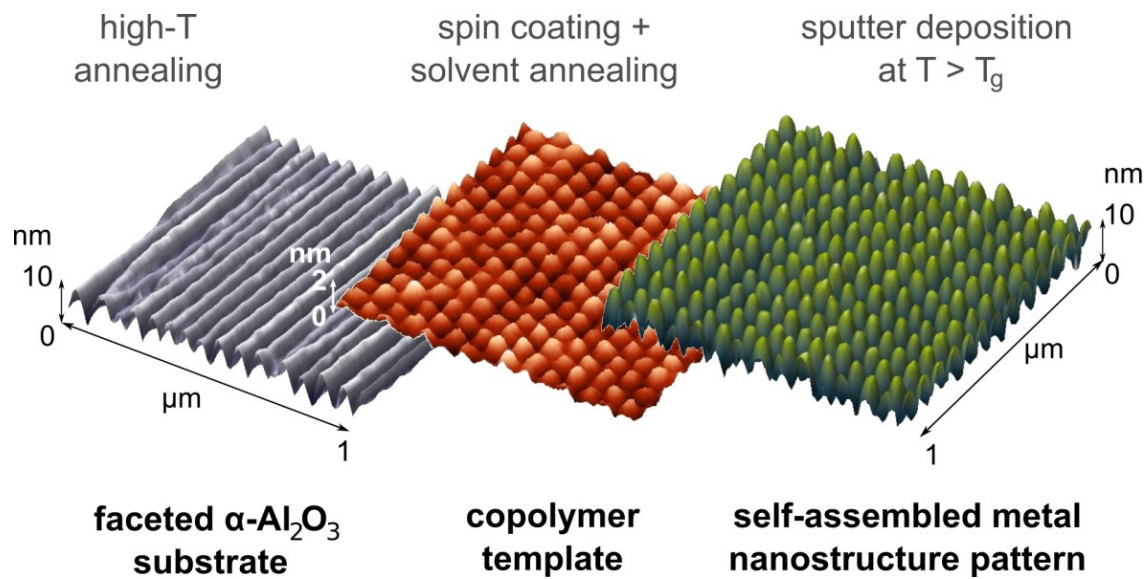

SYNOPSIS

A new bottom-up routine for large-area patterning with metal nanostructures is proposed, employing exclusively self-assembly processes. In-situ GISAXS studies during growth show highly ordered arrays of monodisperse metal nanostructures and evidence tunability of the nanostructure proportions via metal deposition conditions. This routine is distinguished by versatility, ease of implementation, scalability of the nanopattern, and outstanding morphological quality of the nanopatterns, making applications in ultra-high density magnetic data storage or mass production of functional nanostructures conceivable.

ABSTRACT

In a hierarchical nanopatterning routine relying exclusively on self-assembly processes we combine crystal surface reconstruction, microphase separation of copolymers, and selective metal diffusion to produce monodisperse metal nanostructures in highly regular arrays covering areas of square centimeters. In-situ GISAXS during Fe nanostructure formation evidences the outstanding structural order in the self-assembling system and hints at possibilities of sculpting nanostructures by external process parameters. Thus, we demonstrate that nanopatterning via self-assembly is a competitive alternative to lithography-based routines, achieving comparable pattern regularity, feature size, and patterned areas with considerably reduced effort. The option for in-situ investigations during pattern formation, the possibility of customizing the nanostructure morphology, the capacity to pattern arbitrarily large areas with ultra-high structure densities, and the potential of addressing the nanostructures individually enable numerous applications, e.g., in high-density magnetic data storage, in functional nanostructured materials, e.g., for photonics or catalysis, or in sensing based on surface plasmon resonances.

INTRODUCTION

In self-assembly, the size and shape of the resulting structures and the periodicity of structure patterns depend only on the inherent properties and internal interactions of the material which undergoes self-assembly – not on the capability of the tools which the material is processed with. Furthermore, self-assembly is a parallel process – all the individual nanostructures in a pattern form simultaneously, thus preparation can be extremely fast and fabrication time is independent of the sample size[1, 2]. Moreover, there is one crucial advantage of nanostructure fabrication by self-assembly, which is of particular interest from a fundamental scientific point of view: The nanostructures can be studied in-situ during their formation, e.g. regarding size- or shape-dependent characteristics or arising interactions and emerging collective phenomena. Among the diverse self-assembly processes, the microphase separation of block copolymers has attracted much attention due to the morphological variety of these systems. Diblock copolymer self-assembly has been successfully integrated into different nanopatterning and nanostructure fabrication routines. Numerous publications describe routines in which diblock copolymer films fulfill the function of masks, requiring removal of one copolymer block or subsequent removal of both blocks[3-7]. Others demonstrate how selectivity toward the chemical components of a copolymer template can be used to assemble pre-synthesized nanoparticles[8,9]. Only few efforts are made to develop nanopatterning procedures based on the selective wetting of metals deposited on diblock copolymer thin films. Initial works investigated mainly Au and Ag on PS-b-PMMA[10-12]. They first demonstrated the principal possibility of growing self-assembling metal nanostructures on diblock copolymer templates, and achieved granular nanostructures of non-uniform size and rather irregular shape. One recent publication indicates selective wetting of Co sputter deposited onto a PS-b-PEO film, but does neither provide a quantitative analysis of the nanostructure shapes nor investigate effects of the deposition conditions[13]. None of the former studies was concerned with the long-range positional ordering of the metal nanostructures. With our contribution we aim to establish the fabrication of highly-ordered large-area arrays of monodisperse metal nanostructures using exclusively self-assembly processes. We show how the morphology of the metal nanostructure pattern can be adjusted via molecular weight and composition of the diblock copolymer template. We demonstrate that uniform metal nanostructures with well-defined geometric shapes and smooth surfaces can be grown in a facile way by sputter deposition, and that readily accessible process parameters, such as the template temperature during metal deposition, can be utilized to influence the proportions of the metal nanostructures. The nanostructure fabrication routine which we bring forward here is a sequence of three self-assembly processes: 1) the spontaneous reconstruction of $\alpha$-$Al_2O_3$ M-plane surfaces into nanoscale facets[14-16], 2) the microphase separation in diblock copolymer thin films[17-20], and 3) the growth of metal nanostructures on the chemically patterned surface of a diblock copolymer film[11,21]. Our approach is hierarchical in that the pattern formed in one self-assembly process directs the pattern formation in the following process (Fig. 1): First, an $\alpha$-$Al_2O_3$ M-plane substrate is subject to high-temperature annealing, resulting in nanoscale faceting of the substrate surface. In the next step, a diblock copolymer thin film is spin coated onto the nanofaceted $\alpha$-$Al_2O_3$ substrate. Upon exposure to solvent vapor, the copolymer film undergoes microphase separation into uniformly shaped and evenly spaced domains of nanoscale size, consisting exclusively of one of the two copolymer blocks. Here, the $\alpha$-$Al_2O_3$ substrate surface topography defines a pronounced preferential direction, inducing long-range order in the lateral positioning of the chemical domains within the diblock copolymer film[22,23]. Thus, the copolymer film presents a topographically almost flat but distinctly chemically structured surface of high regularity, which can be used as a template to grow a metal nanostructure pattern of the same morphology: Metals exhibiting a pronounced mobility contrast for surface diffusion on the two copolymer blocks will agglomerate on only one of the copolymer blocks[21]. Moreover, shape equilibration to reduce interface energy proceeds significantly slower for nanometer-scale objects than for μm-sized structures[24]. Thus, metal nanostructures can be quasi stable in non-equilibrium shapes (e.g. wires) following the chemical surface pattering of the diblock copolymer template[11, 12]. Thereby the well-ordered two-dimensional chemical structuring of the template surface is transferred into a regular pattern of uniform three-dimensional metal nanostructures.

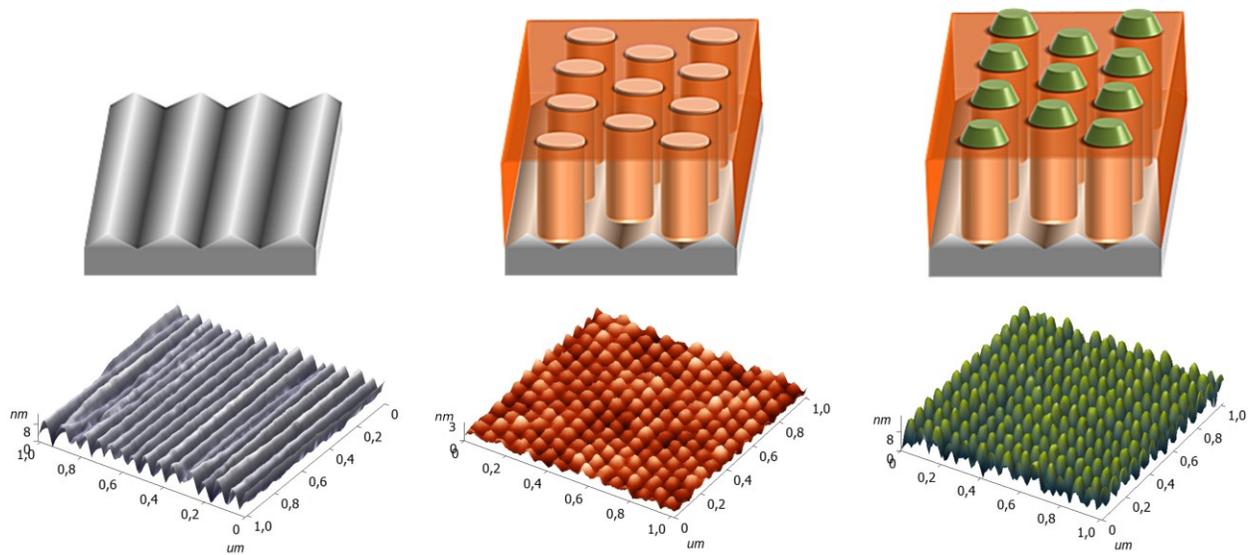

**Figure 1.** Top: Sketch of the proposed nanostructure fabrication routine. Bottom: AFM topography scans of a nanostructures sample in subsequent stages. From left to right: nanofaceted α-Al$_2$O$_3$ substrate; microphase-separated diblock copolymer template (red), the lateral positioning of the chemical domains guided by the substrate topography; metal nanostructures (green) formed during sputter deposition, following the chemical surface patterning presented by the diblock copolymer template.

EXPERIMENTAL

Wafers of polished M-plane α-$Al_2O_3$ were purchased from CrysTec GmbH, cleaned for 15 minutes in an ultrasonic bath of acetone at 50°C, and annealed in air in a chamber furnace at 1325 to 1550 °C for 8 to 48 hours. Lamellae-forming (symmetric) and cylinder-forming (asymmetric) PS-b-PMMA diblock copolymers were purchased from Polymer Standards Service and Polymer Source, Inc. (see Supporting Information, Table 1). The diblock copolymers were dissolved in toluene and spin coated onto the α-$Al_2O_3$ substrates. The copolymer thin films were then dried in air. Microphase separation was accomplished by solvent annealing in saturated acetone vapor at room temperature for 2 h to 3 h. Metal nanostructures were grown on the copolymer templates by magnetron sputter deposition, employing a custom-made UHV chamber at constant template temperatures ranging from room temperature to approximately 200 °C. AFM topography micrographs were recorded ex-situ with an NT-MDT Solver NEXT scanning probe microscope. For in-situ GISAXS experiments performed at the synchrotron x-ray source PETRA III, the sputter deposition chamber was installed at the beamline P01 and measurements were carried out at a wavelength of 14.4 keV. GISAXS patterns were recorded using a MAR345 image plate detector.

RESULTS AND DISCUSSION

The AFM topography micrographs in Fig. 2 compare the surface morphologies of PS-b-PMMA copolymer templates on polished planar silicon substrates with native oxide layer (subfigures 2a and 2c) to those of templates on facetted α-$Al_2O_3$ substrates (subfigures 2b and 2d, the facet edges run from top to bottom of the images). The effect of the substrate topography on the lateral positional ordering of copolymer domains is demonstrated by two types of copolymer templates: a symmetric diblock copolymer forming lamellar domains (BCP-L1) and an asymmetric diblock copolymer forming cylindrical PS domains in a PMMA matrix (BCP-C2). The height differences between the PS and PMMA domains are on average less than 2 nm. For the symmetric diblock copolymer BCP-L1 on polished silicon substrates we observed PS and PMMA alternating in meandering lamellar domains with a spacing of D = 50 nm (peak to peak). The asymmetric diblock copolymer BCP-C2 forms cylindrical domains of PS with a spacing of D = 83 nm (center to center) in a PMMA matrix. Orientation of the domains perpendicular to the film interfaces is ensured by film thicknesses d < 0.5 D [25]. Although the polymer domains are very uniform in size, shape, and orientation, there is only short-range order in their lateral arrangement. On the facetted α-$Al_2O_3$ substrates on the other hand, long-range positional ordering is obtained: For the symmetric diblock copolymer BCP-L1, all lamellae align parallel to the substrate facets. The fast Fourier transform (FFT) of the topography shows sharp maxima at positions corresponding to the translational symmetry and lateral spacing of the chemical domains. For the asymmetric diblock copolymer BCP-C2, the PS domains are arranged in a 2D hexagonal lattice with the orientation of the (10) axis given by the orientation of the substrate facet edges. The respective FFT exhibits the six-fold symmetry of the hexagonal domain array with peak positions corresponding to the domain spacing. In both cases, the long-range precision of the lateral domain positioning is evidenced by the higher order maxima in the FFTs.

Different metals were sputter deposited onto diblock copolymer templates at various temperatures and deposition rates. For the noble metal Au we observed the formation of small metal clusters with their positioning rather weakly influenced by the chemical surface pattering of the template for most deposition conditions. The base metals Fe, Pt, and Ni, however, readily formed uniform and well-separated nanostructures on the PS domains, with sizes, shapes, and arrangement given by the chemical surface pattern of the templates (see Figure 3). The disparate behavior of Au and the base metals is attributed to a different balance of metal/metal and metal/polymer interactions[11, 12, 21]. Given that a metal reproduces the template pattern well, control over shape, size, and lateral arrangement of the metal nanostructures is obtained via control over the morphology of the diblock copolymer template: The shapes of the metal nanostructures follow the copolymer domain morphology at the film surface, as given by the volume fractions of the copolymer blocks. The metal nanostructures can further be scaled in size by providing templates with desired sizes of the copolymer domains (ranging from less than 5 nm to a few hundred nm), as determined by the molecular mass of the diblock copolymer (compare Fig 4 b and d). Dot-shaped or wire-like metal nanostructures with a high degree of regularity in lateral positioning can be prepared on diblock copolymer templates with cylindrical or lamellar domain morphology, respectively, on nanofaceted α-$Al_2O_3$ substrates, as shown in Fig 4 (see also Supporting Information for AFM topographies of larger sample areas). Thus, the periodic substrate topography guides the chemical surface patterning of the diblock copolymer template, which is in turn transferred into a highly regular pattern of three-dimensional metallic nanostructures. The sample sizes with nanostructure patterns covering areas of up to 3 $cm^2$ were chosen for ease of sample handling. Since the proposed routine employs exclusively self-assembly processes, the samples can be readily scaled up to much larger sizes, without increasing duration or complexity of the preparation procedure.

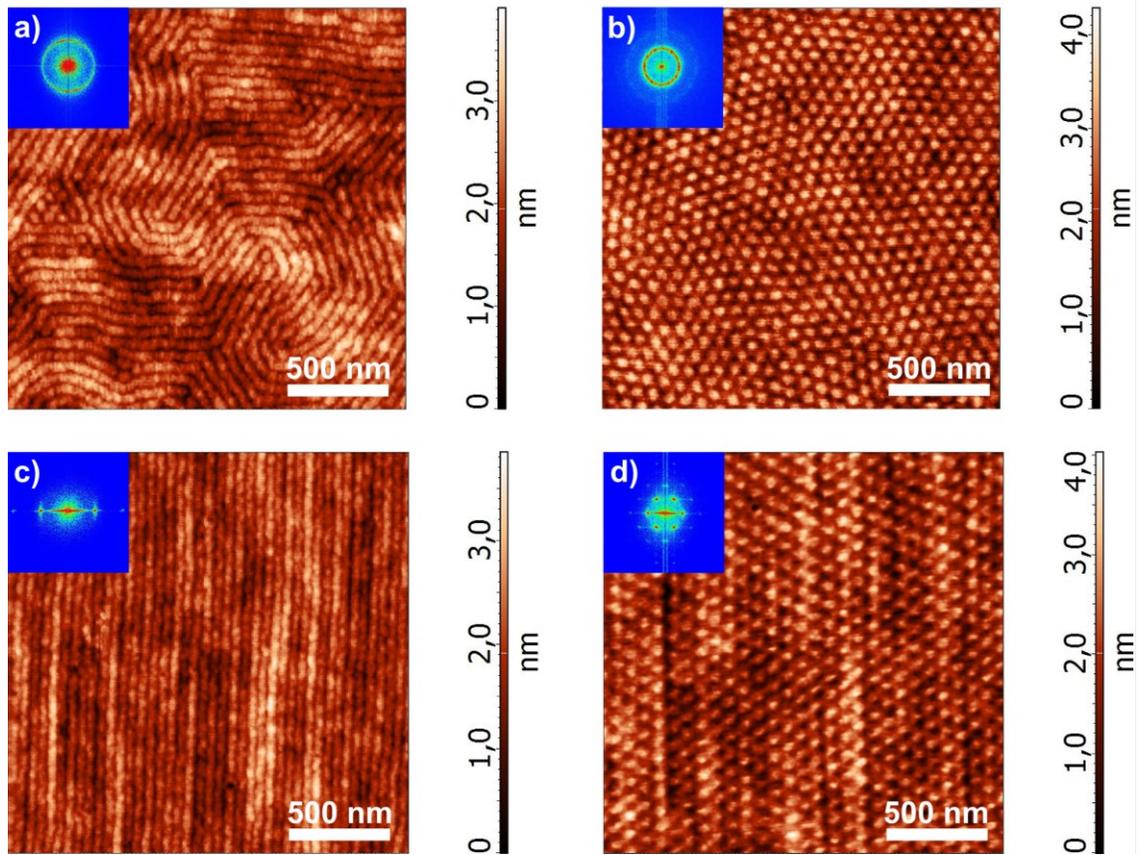

**Figure 2.** AFM topography micrographs comparing the chemical surface patterning of different diblock copolymer thin film templates resulting from microphase separation. PS domains appear bright, PMMA domains appear dark. Fast Fourier transforms of the respective topographies are shown. Thin films of the diblock copolymers BCP-L (lamellar domain morphology) and BCP-C2 (cylindrical domain morphology) were prepared on planar $SiO_x$ substrates (a, b) and on nanofaceted α-$Al_2O_3$ substrates (c, d). The uniaxially corrugated topography of the α-$Al_2O_3$ substrate induces long-range ordering in the lateral positioning of the chemical domains of the diblock copolymer thin film.

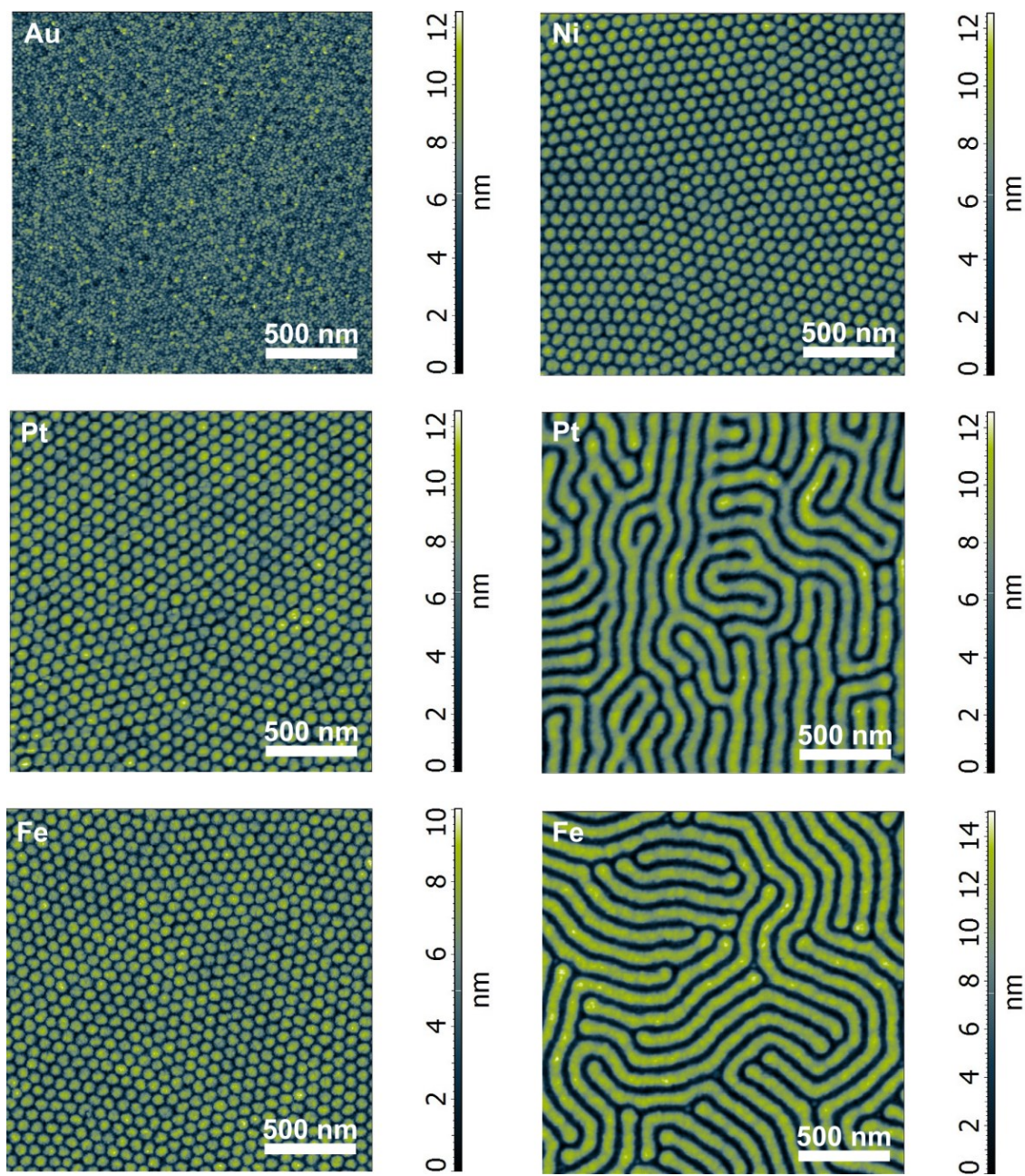

**Figure 3.** Nanostructures of Au, Ni, Fe, and Pt grown on diblock copolymer templates with cylindrical and lamellar morphology (BCP-C2, BCP-L2) on planar $SiO_x$ substrates. While Au forms very small clusters all over the template surface, the other metals form smooth and well-separated nanostructures exactly reproducing the chemical patterning of the template surface.

Due to the temperature dependence of surface diffusion processes and of the time scale for shape equilibration[24], an effect of the template temperature on the resulting nanostructure morphology is expected. AFM micrographs, however, are compromised by the probe size and by convolution of the nanostructure shape with the tip shape, and thus hardly allow for unambiguous determination of the nanostructure shape. SEM measurements would be hindered by the fact that the sample surface is not continuously conductive. Moreover, it is not trivial to realize in-situ microscopy measurements on nanostructures during growth. To observe the size, shape, and lateral positioning of iron nanostructures during their formation at different constant template temperatures, we performed in-situ Grazing Incidence Small Angle X-ray Scattering (GISAXS) during iron sputter deposition. Two hexagonal arrays of iron nanodots were grown on BCP-C2 templates on nanofaceted α-$Al_2O_3$ at room temperature and at approximately 170 °C, respectively. The deposition processes were interrupted regularly to obtain GISAXS patterns.

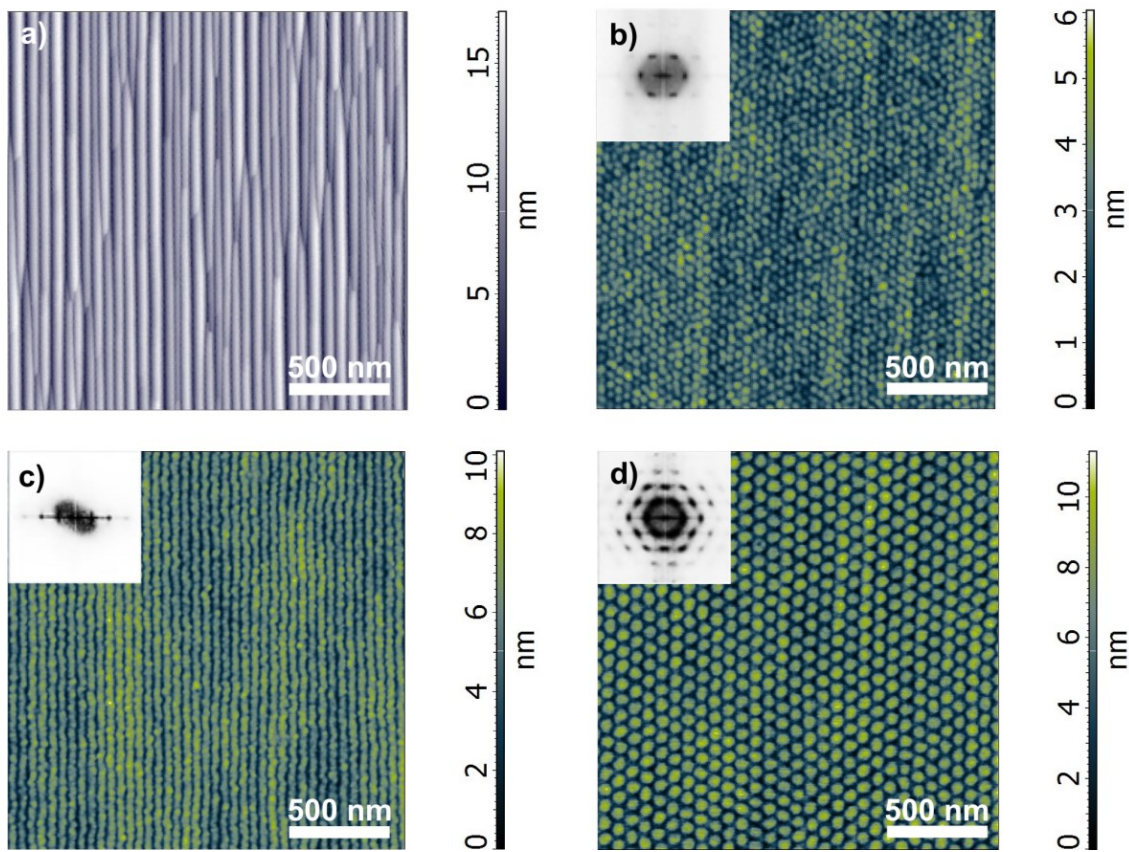

**Figure 4.** Diblock copolymer templates on nanofaceted α-Al₂O₃ substrates (a) exhibit a highly regular chemical surface pattern, on which uniform dot-shaped (b, d) or wire-like (c) nanostructures with highly ordered lateral positioning are grown. Comparing the nanostructures grown on templates of copolymers BCP-C1 and BCP-C2 with identical domain morphology but different domain sizes illustrates how the nanostructure size can be scaled via the copolymer molecular weight. Moreover, the nanostructure pattern can be reversed by reversing the volume fractions of PS and PMMA in the diblock copolymer to produce antidots. See Supporting Information, figure S1, for large-area AFM topography scans.

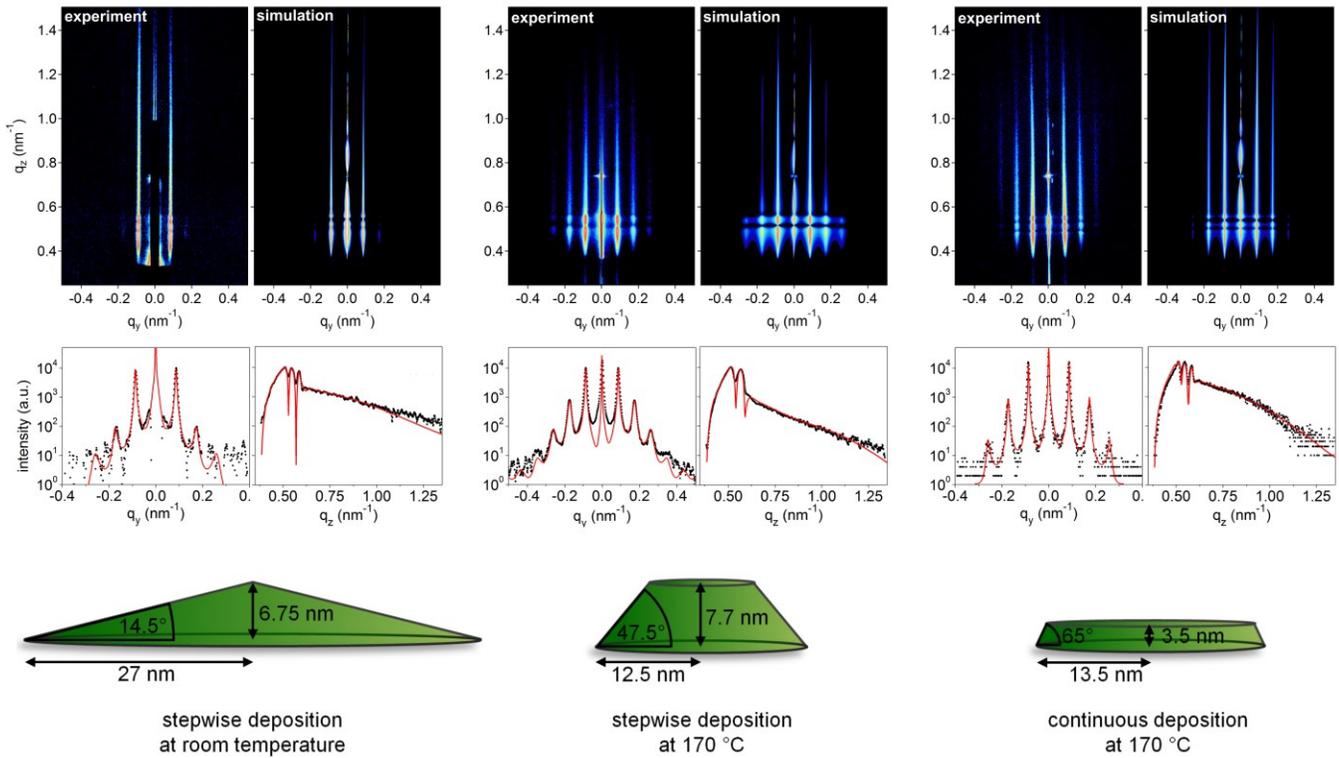

**Figure 5.** Morphologies of self-assembled Fe nanodots in hexagonal arrays forming under different metal deposition conditions. Top: GISAXS scattering pattern and corresponding 2D simulations. Middle: Horizontal and vertical sections through the experimental and simulated scattering patterns as indicated above. Bottom: The sketches depict the resulting proportions of the nanodots as deduced from the simulations.

GISAXS data were evaluated using the software IsGISAXS[26], yielding values for the lateral nanostructure arrangement and for the development of the nanostructure dimensions during growth. In choosing the according parameters for the simulation, priority was given to the assumption of a linearly increasing nanodot volume with $V(t_0) = 0$ and to an optimum fit in the intensity distribution $I(q_y)$. Fig. 5 shows exemplary experimental data with corresponding simulations for both iron nanodot arrays (please refer to Supporting Information for data and simulations for all deposition stages). Analogous information is given for one nanodot array grown in continuous deposition at approximately 170°C without interruptions for scattering data acquisition; the GISAXS pattern was recorded at the end of the deposition process.

The fact that no distributions in particle dimensions or arrangement had to be assumed for the simulations, demonstrates the outstanding morphological uniformity and positional regularity of nanostructure arrays prepared via the proposed routine. All scattering patterns evidence a hexagonal arrangement of nanodots with a lattice constant of 83 nm as given by the diblock copolymer template. Among the form factors implemented in the simulation program, truncated cones provided the best possible approximation to the average nanodot shape. Given the high degree of uniformity of the nanodots, it is valid to consider this average particle shape to correspond well to the shape of an individual nanodot. However, no satisfying agreement between simulation and experimental data could be found for the early stages of the sample grown at 170°C, indicating that here the assumed form factor does not correspond well to the actual nanodot shape. Assuming a significantly larger radius would yield a better fit, but results in an unphysical development of the nanodot volume. Geometrical parameters describing the nanodot shape as extracted from simulations for all stages of growth are summarized in Fig. 6. Under both growth conditions, the base angle and the ratio of height to base radius remain roughly constant during growth, irrespective of the different deposition rates. The quantitative effect of the template temperature on the nanodots, however, is remarkable: At any given deposition stage the aspect ratio of height to base radius is about 0.65 for the nanodots growing at 170 °C, and merely about 0.25 for the nanodots growing at room temperature. The base angle is more than three times as large for the nanodots growing at 170 °C as compared to the sample at room temperature; for continuous Fe deposition the angle is even more than four times larger. In the sample grown at room temperature, the nanodots eventually extend their radius over the radius of the PS domains of approx. 29 nm, rather than growing more in height. Apparently, the nanodot proportions strongly depend on the template temperature due to its effect on the iron atoms' kinetics on the template surface: At room temperature, the equilibration time of nm-scaled bodies can easily reach the range of months[21, 24]. Consequently, surface area minimization of the nanodots is impeded and growth at room temperature results in wide, flat nanodots. At higher template temperatures the mobility of iron atoms and clusters is enhanced, and the equilibration time can be reduced to the range of a few hours. Thus, compact iron nanodots with lower surface to volume ratio form. The very different nanodot proportions resulting from different growth conditions (Fig. 5) hint at an approach to shaping self-assembling nanostructures by controlling external process conditions.

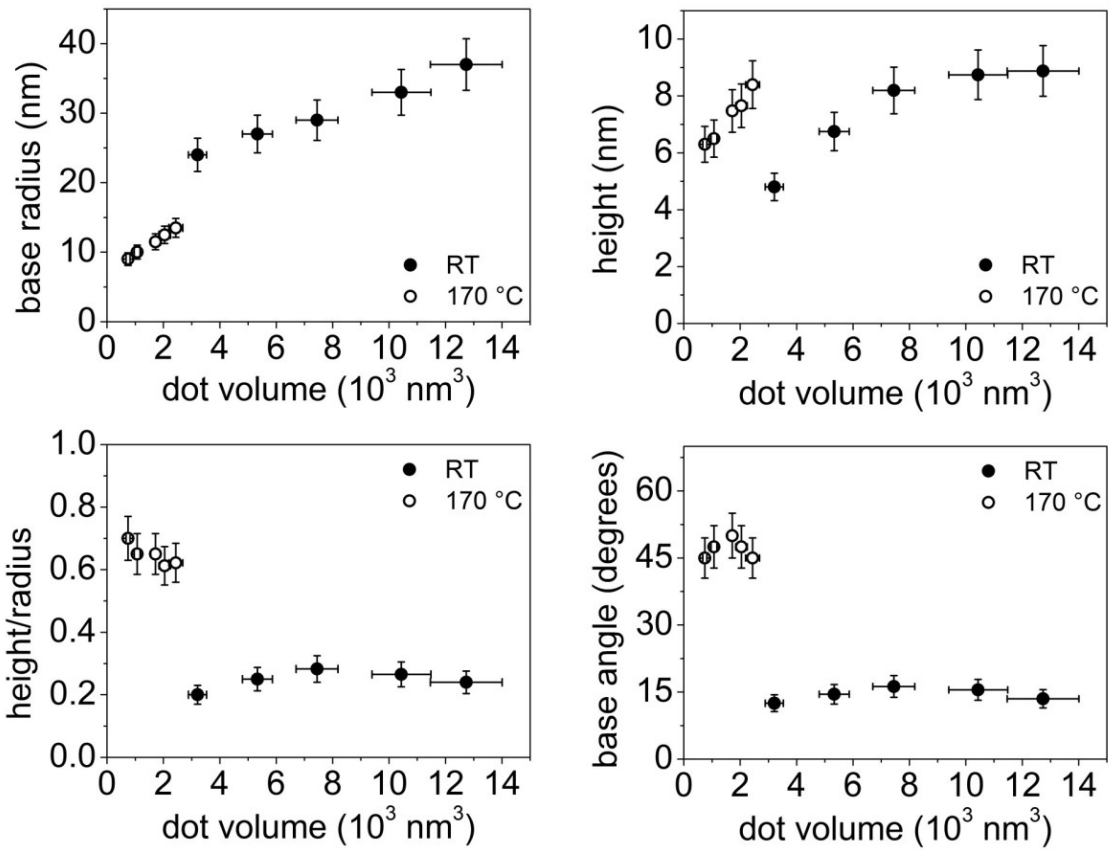

**Figure 6.** Evolution of the geometrical parameters describing the proportions of Fe nanodots forming at 170 °C and at room temperature, respectively, as extracted from simulations of in-situ GISAXS data (see Supporting Information, figures S2, S3).

CONCLUSION

Our novel routine brings two approaches in nanostructure fabrication together: The directed microphase separation of diblock copolymer films on nanostructured substrate surfaces was combined with the self-assembly of metal atoms on microphase-separated diblock copolymer templates. Employing exclusively self-assembly processes eliminates the requirement for complex devices for lithographical nanopatterning and allows for in-situ studies throughout the entire procedure. Following our hierarchical self-assembly routine, diverse metal nanostructure patterns with a high degree of morphological uniformity and positional regularity can be prepared on large sample areas. Our findings inferred from in-situ GISAXS during Fe nanodot growth indicate that the proportions of nanostructures can be influenced via deposition conditions such as the template temperature. The outstanding monodispersity of the nanostructures should allow for identifying interactions among them and for studying nanostructure properties accurately without the need to isolate individual nanostructures from the ensemble. The morphological quality of the nanostructures also encourages extending the nanopatterning to morphologies provided by triblock copolymers: These comprise, among many others, morphologies with cubic or kagome patterning, which are of great interest for studies of magnetic frustration. The proposed routine is fast, facile, versatile and economic. It offers exciting possibilities: covering arbitrarily large surface areas with various patterns of customizable identical nanostructures, achieving ultra-high structure densities with nanostructure diameters below 5 nm, or addressing individual nanostructures in a highly-ordered array. High-density magnetic recording, surface plasmon resonance based sensing, or materials for catalysis may be among the future applications of these self-assembling nanopatterned composite systems.

ACKNOWLEDGMENT

The authors thank Hans-Christian Wille and Frank-Uwe Dill at the beamline P01 at PETRA III for their unfailing support in preparing and conducting the GISAXS experiments.

**Corresponding Author**

\* Denise Erb

Deutsches Elektronen-Synchrotron DESY, Notkestraße 85, D-22607 Hamburg, Germany

denise.erb@desy.de

**Present Address**

† Denise Erb

The Hamburg Centre for Ultrafast Imaging CUI, Luruper Chaussee 149, 22761 Hamburg, Germany

denise.erb@uni-hamburg.de

SUPPORTING INFORMATION

| reference name | domain morphology | total molecular weight | volume fraction of PS | poly-dispersity | equilibrium domain period |
|---|---|---|---|---|---|
| BCP-L1 | lamellar | 100 kg/mol | 47 % | 1.12 | 48 nm |
| BCP-L2 | lamellar | 406 kg/mol | 50 % | 1.10 | 103 nm |
| BCP-C1 | cylindrical | 94 kg/mol | 28 % | 1.18 | 48 nm |
| BCP-C2 | cylindrical | 205 kg/mol | 31 % | 1.08 | 83 nm |

**Table 1.** Properties of the employed diblock copolymers.

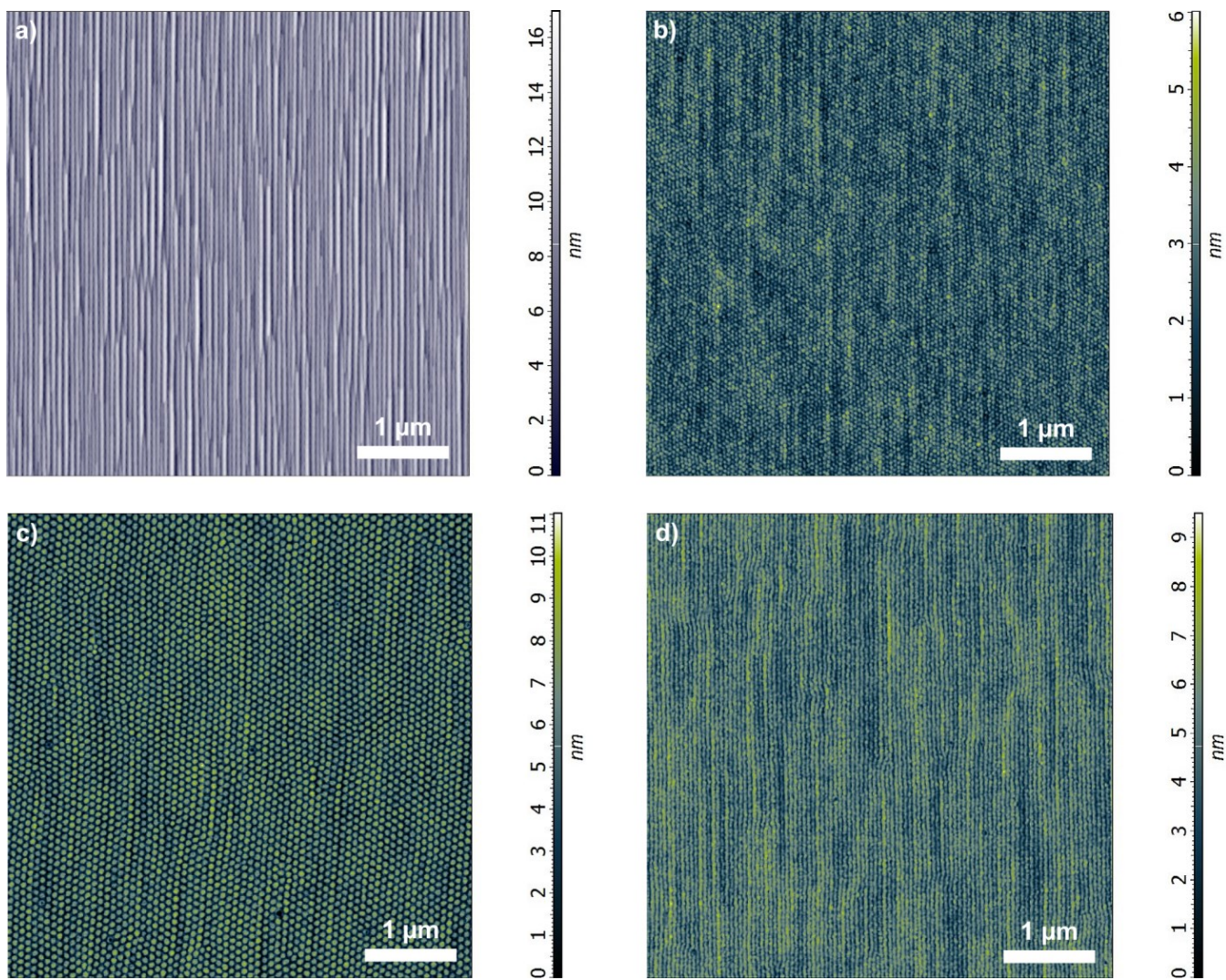

**Figure S1.** Large-area AFM micrographs of a) nanofaceted substrate surface and b) - d) Fe nanostructure patterns grown on symmetric and asymmetric diblock copolymer templates. Long range lateral ordering is induced by the substrate; the different nanostructure patterns illustrate some of the morphological options and the scalability of the domain size of the diblock copolymer templates.

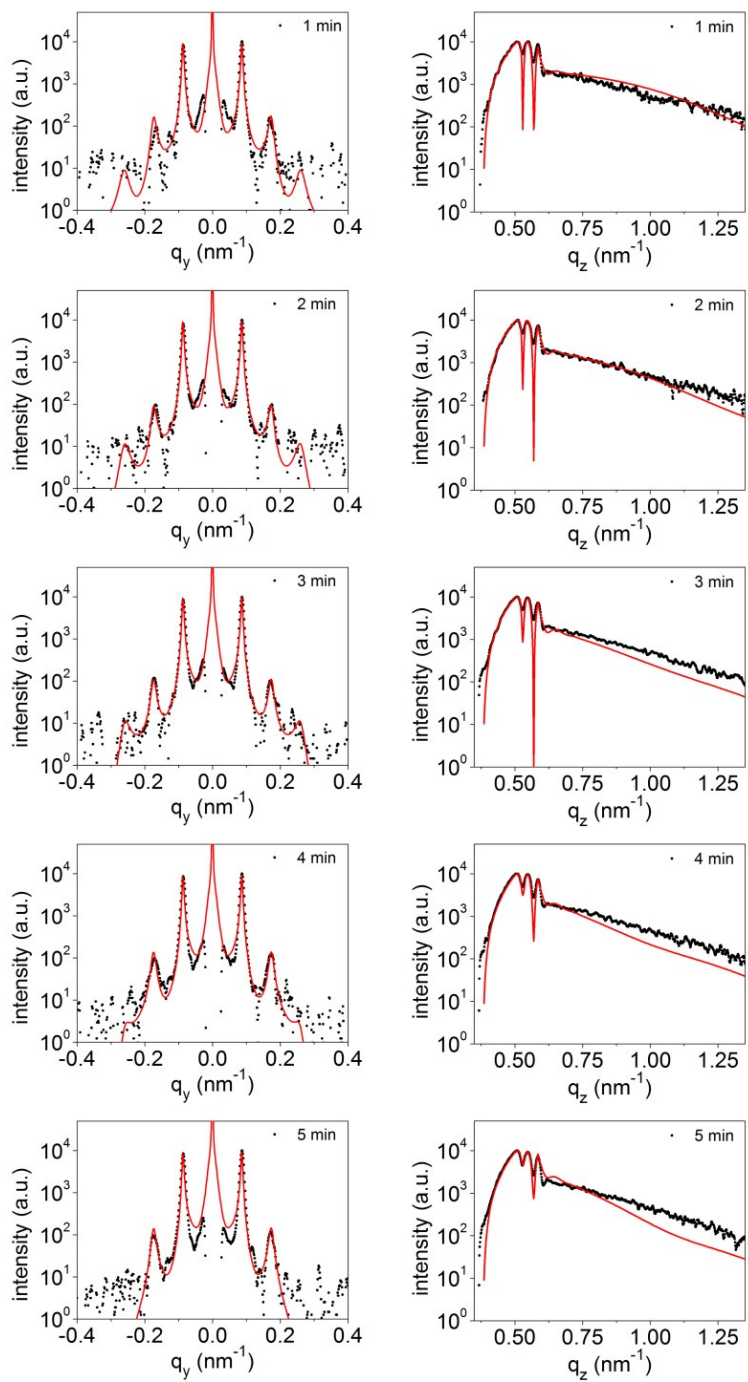

**Figure S2.** Sections in qy and qz direction through a sequence of GISAXS patterns, recorded in-situ during Fe nanodot growth at room temperature, with simulations (red solid lines). Labels indicate the elapsed Fe deposition time.

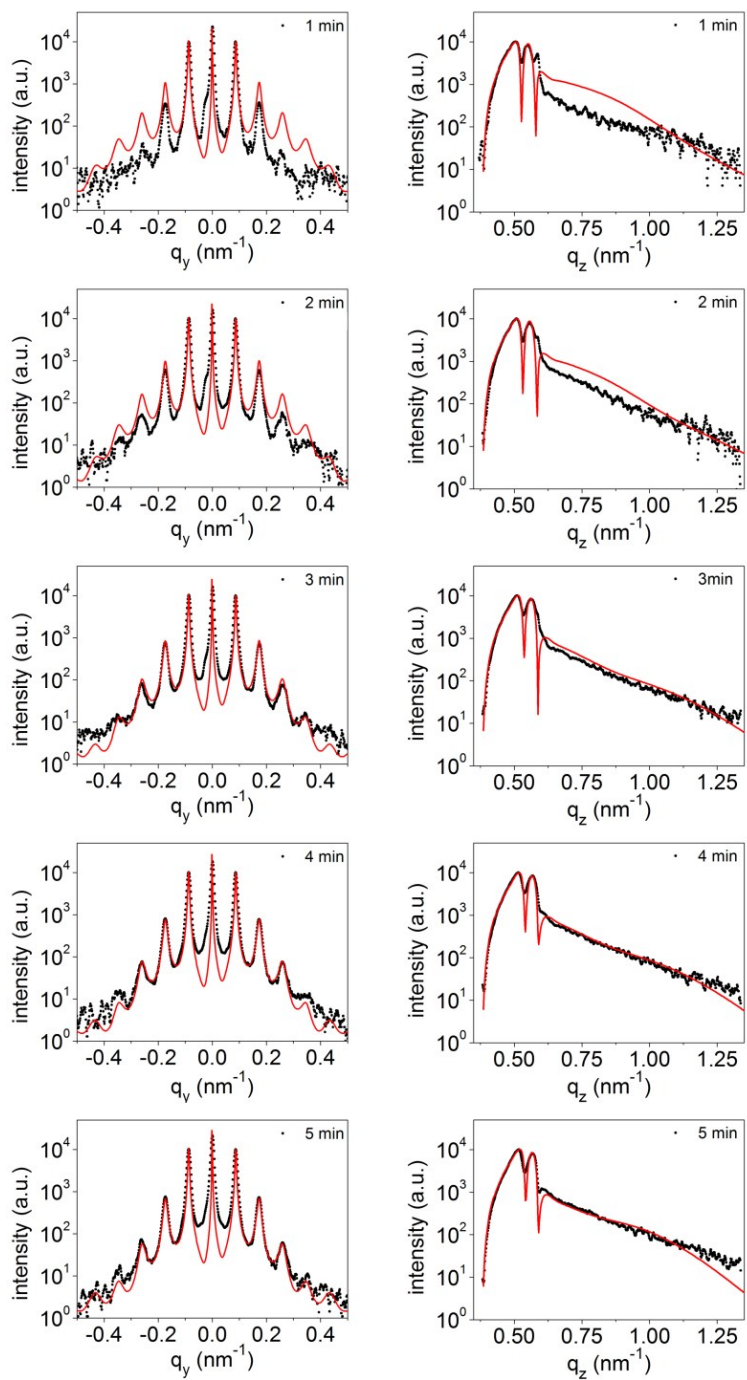

**Figure S3.** Sections in qy and qz direction through a sequence of GISAXS patterns, recorded in-situ during Fe nanodot growth at 170 °C, with simulations (red solid lines). Labels indicate the elapsed Fe deposition time.